\documentclass[aps,amssymb,prb,twocolumn]{revtex4}
\usepackage{graphicx}
\usepackage{epsfig,amsmath}

\begin{document}

\title{Thermal rectification in nonlinear quantum circuits}
\author{Tomi Ruokola$^1$}
\email[Correspondence to ]{tomi.ruokola@tkk.fi}
\author{Teemu Ojanen$^2$}
\email[Correspondence to ]{ojanen@physik.fu-berlin.de}
\author{Antti-Pekka Jauho$^{1,3}$}
\affiliation{$^1$ Department of Applied
Physics, Helsinki University of Technology, P.O.~Box 1100,
FIN-02015 HUT, Finland }
\affiliation{$^2$ Institut f\"ur Theoretische Physik, Freie Universit\"at Berlin,
Arnimallee 14, 14195 Berlin, Germany}
\affiliation{$^3$ Department of Micro and Nanotechnology,
Technical University of Denmark, Building 345 East,
DK-2800 Kongens Lyngby, Denmark }

\date{\today}
\begin{abstract}

We present a theoretical study of radiative heat transport in
nonlinear solid-state quantum circuits. We give a detailed account of
heat rectification effects, i.e. the asymmetry of heat current with
respect to a reversal of the thermal gradient, in a system consisting
of two reservoirs at finite temperatures coupled through a nonlinear
resonator. We suggest an experimentally feasible superconducting
circuit employing the Josephson nonlinearity to realize a controllable
low temperature heat rectifier with a maximal asymmetry of the order
of 10\%. We also discover a parameter regime where the rectification
changes sign as a function of temperature.

\end{abstract}
\pacs{PACS numbers: } \bigskip

\maketitle

\section{Introduction}
Heat transport in nanoscale structures has become an active and
rapidly growing research area. Progress in experimental methods has
enabled the study of fundamental issues, and lately the field has
seen major breakthroughs, such as the measurement of quantized heat
transport, \cite{schwab} and manipulation of thermal currents using
external control fields. \cite{meschke, saira} In solid-state
systems electron--electron and electron--phonon scattering are the
most important channels for small systems to exchange energy with
the environment. However, recently it was understood that at low
temperatures one needs to take into account the radiative channel
which becomes the dominant relaxation method in mesoscopic samples
below the phonon--photon crossover. \cite{meschke,schmidt,ojanen}

\begin{figure}[ht]
\centering
\includegraphics[width=\columnwidth]{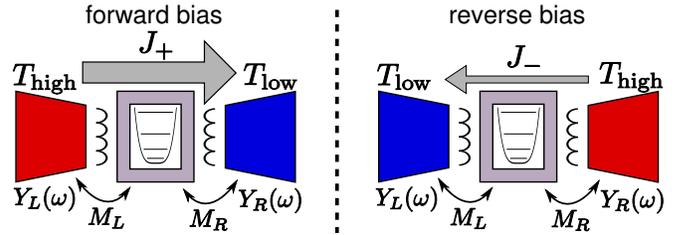}
\caption{In a heat transport experiment, thermal energy flows from a
hot reservoir (temperature $T_{\rm high}$) to a cold reservoir (temperature
$T_{\rm low}$). To obtain the  rectification $\cal R$ one must measure the
current for both thermal bias directions. In our model the heat is
transported by inductive magnetic coupling between the reservoirs
and the central nonlinear resonator.}\label{fig1}
\end{figure}

In this paper we study rectification effects in
thermal transport mediated by electromagnetic
fluctuations in solid-state nanostructures.
In a two-terminal geometry a finite
rectification means that heat current is not simply reversed when
the thermal gradient changes sign, but also the absolute magnitude
of the current changes. We define the rectification $\cal R$ as
\begin{equation}
{\cal R} = (J_+-J_-)/\max\{J_+,\,J_-\}\label{recti},
\end{equation}
where $J_+$ and $J_-$ are the magnitudes of the heat currents in
forward and reverse bias configurations, respectively (see Fig.~\ref{fig1}).
Previously rectification has been shown to take place in systems where
a classical \cite{terraneo,li,hu} or quantized \cite{segal2,zeng}
nonlinear chain is coupled asymmetrically to linear reservoirs,
when nonlinear reservoirs are coupled through a harmonic oscillator,
\cite{segal} or in hybrid quantum junctions.\cite{wu}
  Here we demonstrate rectification in a fully
quantum-mechanical and experimentally realizable model where
photon-mediated heat current flows between two linear reservoirs
coupled asymmetrically to a nonlinear resonator.

Our analysis is based on a nonequilibrium Green's function method
developed in Ref.~\onlinecite{ojanen2}, and the nonlinear transport problem
is solved with a self-consistent Hartree approximation.
Rectification is studied as a function of the operating temperatures,
reservoir coupling strengths and admittances, and the strength of the
nonlinearity. We also propose a concrete setup based on a
superconducting quantum interference device (SQUID) where the
rectification effects can be realized with current experimental
technology at sub-Kelvin temperatures. A similar circuit, operated in
the linear regime, was employed in the pioneering experiment
demonstrating photonic heat transport. \cite{meschke} By adjusting the
external magnetic flux through the circuit it is possible to tune the
rectification continuously between zero and the maximum value. Using
realistic parameters we find a rectification of over 10\%, and
identify a regime where $\cal R$ changes sign as a function
of temperature.  Experimentally rectification has been observed in
phonon transport through a nanotube\cite{chang}  at room temperature
with ${\cal R} = 7\%$ and in electron transport through a
quantum dot\cite{scheibner} at
80~mK with $\cal R$ up to 10\%. 

\section{Model}

The thermal transport setup is depicted in Fig.~\ref{fig1}. It consists of
two linear reservoir circuits with admittances $Y_L(\omega)$ and
$Y_R(\omega)$. Temperatures of the left and right reservoirs are
$T_{\rm high}$ and $T_{\rm low}<T_{\rm high}$ in the forward bias setting, and vice versa for
reverse bias. We assume that heat can flow between the reservoirs
only through a mediating nonlinear resonator circuit. The couplings
between the reservoirs and the resonator are taken to be inductive
with mutual inductances $M_L$ and $M_R$. Using the Caldeira--Leggett
mapping between linear admittances and bosonic reservoir modes the
total Hamiltonian takes the form $H=H_L+H_R+H_{M}+H_{C}$, where the
middle circuit and reservoir terms are
\begin{align}\label{nonlin}
H_M=\hbar\omega_0(\hat{b}^{\dagger}\hat{b}+\frac{1}{2})+
\frac{\hbar\epsilon}{2}(\hat{b}+\hat{b}^{\dagger})^4,
\end{align}
\begin{align}
H_{L/R}=\sum_{j\in
L/R}\hbar\omega_j(\hat{a}_j^{\dagger}\hat{a}_j+\frac1 2),
\end{align}
and the inductive coupling term is
\begin{align}\label{coupl}
H_{C}=\hat{I}\left(M_L\hat{i}_{L}+ M_R\hat{i}_{R}\right),
\end{align}
which involves the current operators for the central device $\hat{I}$
and for the reservoirs $\hat{i}_{L/R}=\sum_{j\in
L/R}g_j(\hat{a}_j+\hat{a}_j^{\dagger})$, respectively. The
 electric current operator for the central
device can be expressed as
$\hat{I}=I_0(\hat{b}+\hat{b}^{\dagger})$ with
$I_0=\sqrt{\hbar\omega_0/2L}$ and $\omega_0=1/\sqrt{LC}$ where $L$ and
$C$ are the linear inductance and capacitance of the resonator;
$\hat{b}$, $\hat{b}^{\dagger}$ and reservoir operators are bosonic
creation and annihilation operators,
$[\hat{b},\hat{b}^{\dagger}]=1$. The nonlinearity of the central
circuit is characterized by the the second term in Eq.~(\ref{nonlin}),
corresponding to a quartic potential whose strength is controlled
by the parameter $\epsilon$.
It must be emphasized that Eq.~(\ref{coupl}) has a
generic bilinear form and therefore our results
are relevant for other types of systems beyond the studied
realization.

The basis of our analysis is provided by the Meir--Wingreen formula
for the heat current \cite{ojanen2,remark}
\begin{align}\label{current4}
J=&\int_{0}^{\infty}\frac{d\omega\omega^2M_L^2}{2\pi} \left\{
2\left[S_I(\omega)-S_I(-\omega)\right]\mathrm{Re}[Y_{L}(\omega)]n_L(\omega)\right.\nonumber\\
&\left.{}-2S_I(-\omega)\mathrm{Re}[Y_{L}(\omega)]\right\}.
\end{align}
Here $n_L(\omega)$ is the Bose function of the left reservoir  and
$S_I(\omega)=\int_{-\infty}^{\infty}dt\,e^{i\omega(t-t')}
\langle\hat{I}(t)\hat{I}(t')\rangle$
is the current noise power of the central circuit.
The admittances $Y_{L/R}(\omega)$ are related to the current
correlation functions of the free reservoirs. \cite{ojanen2} In the
absence of the nonlinear term $(\epsilon=0)$ the transport problem
can be solved exactly for arbitrary couplings and reservoir
admittances. \cite{ojanen2} No rectification takes place in this
regime. In the following we solve the nonlinear transport problem in
a self-consistent Hartree approximation, which is expected to be
accurate for small values of the nonlinearity. This approach does
not fully account for the correlation effects due to the interplay
of nonlinearity and tunneling which are potentially important in the
ultra-low temperature regime $T_{\rm high},\,T_{\rm low}\ll \hbar\omega_0/k_B$.
However, analogously to interacting electron transport problems, the
mean-field approach is accurate in the sequential tunneling regime
when the temperatures are of the order of $\hbar\omega_0/k_B$.

As a first step we
approximate the resonator Hamiltonian as
\begin{equation}\label{int1}
H_M\approx\hbar\omega_0(\hat{b}^{\dagger}\hat{b}+\frac{1}{2})
+3\hbar\epsilon\Phi(\hat{b}^{\dagger}+\hat{b})^2,
\end{equation}
where we have used $(\hat{b}^{\dagger}+\hat{b})^4\approx
6\Phi(\hat{b}^{\dagger}+\hat{b})^2$ with the mean field
$\Phi=\langle(\hat{b}^{\dagger}+\hat{b})^2\rangle$.
Here the factor 6 is the number ways two operators can be picked
from a set of four.
By performing a diagrammatic expansion of the resonator Green's
function one can show that this procedure is identical to
the self-consistent Hartree approximation.
Because
Eq.~(\ref{int1}) is now quadratic in bosonic operators, it is
possible to bring it to diagonal form by a canonical
transformation. However, now we have the added complication of an a
priori unknown mean field, which has to be evaluated
self-consistently in a nonequilibrium state. The transformed
Hamiltonian and current operators are
\begin{align}\label{mf}
&H_{M} = \hbar\tilde{\omega}_0(\tilde{b}^\dagger\tilde{b}+\frac 1 2),
\qquad\hat{I}=\tilde{I}_0(\tilde{b}+\tilde{b}^\dagger),
\end{align}
where
$\tilde{\omega}_0=\omega_0\sqrt{1+\frac{12\epsilon\Phi}{\omega_0}}$ and
$\tilde{I}_0 = \sqrt{\frac{\omega_0}{\tilde{\omega}_0}}I_0$. Thus the
effect of the nonlinear term is incorporated by a mean-field dependent
renormalization of the resonance frequency of the oscillator and its
current operator. For further development it is convenient to
introduce the correlation functions
$\langle\hat{I}(t)\hat{I}(t')\rangle^{r}=-i\theta(t-t')\langle[\hat{I}(t),\hat{I}(t')]\rangle$
and $\langle\hat{I}(t)\hat{I}(t')\rangle^{<}=
-i\langle\hat{I}(t')\hat{I}(t)\rangle$. A nonequilibrium
equation-of-motion analysis \cite{haug}, similar to the one presented
in Ref.~\onlinecite{ojanen2}, reveals that the current correlators are given
by
\begin{eqnarray}
\langle\hat{I}\hat{I}\rangle^{r}(\omega)&=&
\frac{1}{\left(\langle\hat{I}\hat{I}\rangle^{r}_0(\omega)\right)^{-1}
-\tilde{I}_0^{-2}\Sigma^r(\omega)}, \label{iir} \\
\langle\hat{I}\hat{I}\rangle^{<}(\omega)&=&\tilde{I}_0^{-2}
|\langle\hat{I}\hat{I}\rangle^{r}(\omega)|^2\Sigma^{<}(\omega),\label{iiless}
\end{eqnarray}
where
$\langle\hat{I}\hat{I}\rangle^{r}_0(\omega)=2\tilde{I}_0^2\tilde{\omega}_0/(\omega^2-\tilde{\omega}_0^2)$
is the retarded Green's function of the uncoupled oscillator. The self-energies
\begin{eqnarray}
\Sigma^r(\omega)&=&-\frac{i\tilde{I}_0^2\omega}{\hbar}
\Big[M_L^2Y_L(\omega)+M_R^2Y_R(\omega)\Big], \label{sigmar}\\
\Sigma^{<}(\omega)&=&-\frac{2i\tilde{I}_0^2\omega}{\hbar}
\Big[M_L^2\mathrm{Re}[Y_L(\omega)]n_L(\omega){}\nonumber\\
&&{}+M_R^2\mathrm{Re}[Y_R(\omega)]n_R(\omega)\Big],\label{sigmaless}
\end{eqnarray}
take into account the presence of reservoirs. Furthermore, the mean
field $\Phi$ is related to the lesser correlator via
\begin{align}\label{phi}
&\Phi=\langle(\hat{b}^{\dagger}+\hat{b})^2\rangle=
-I_0^{-2}\int_{-\infty}^{\infty}\frac{d\omega}{2\pi i}\langle\hat{I}\hat{I}\rangle^{<}(\omega).
\end{align}
Equations (\ref{iir})--(\ref{phi}) form a closed set of
equations which needs to be solved to find the current correlation
functions. The self-consistent solution proceeds by making an
initial guess for the mean field, calculating the correlation
function $(\ref{iiless})$ corresponding to the initial value and
calculating the updated value of the mean field by evaluating the
integral in Eq.~(\ref{phi}). The  procedure is repeated until
convergence is achieved. The current noise then follows immediately
from the lesser function
$S_I(\omega)=-\mathrm{Im}\langle\hat{I}\hat{I}\rangle^{<}(-\omega)$
which yields the heat current after evaluating Eq.~(\ref{current4}).
In the case of a vanishing nonlinearity $(\epsilon=0)$ this
procedure recovers the exact solution of the linear problem.
To facilitate the analysis of the rectifying mechanism we note that
with the help of Eqs.~(\ref{iir})--(\ref{phi}) we can write
Eq.~(\ref{current4}) in the form
\begin{align}\label{curr5}
J=\int_{0}^{\infty}\frac{d\omega}{2\pi}
\frac{4\hbar\omega^3M_L^2M_R^2{\rm Re}[Y_L]{\rm Re}[Y_R]
(n_L-n_R)}
{\big|F(\omega)+i\omega[M_L^2Y_L+M_R^2Y_R]\big|^2}
\end{align}
where
$F(\omega)=\hbar(\omega^2-\omega_0^2-12\omega_0\epsilon\Phi)/
(2I_0^2\omega_0)$. 
The frequency dependence of $Y_{L/R}(\omega)$ and $n_{L/R}(\omega)$
has been suppressed for brevity.

For numerical calculations explicit expressions for the admittances
$Y_{L/R}(\omega)$ are needed. Here we assume that the reservoir
circuits effectively consist of a resistor, a capacitor, and an
inductor in series, resulting in
$Y_{L/R}(\omega)=R_{L/R}^{-1}[1-iQ_{L/R}(\frac{\omega}{\omega_{L/R}}-
\frac{\omega_{L/R}}{\omega})]^{-1}$, where $R_{L/R}$, $Q_{L/R}$, and
$\omega_{L/R}$ are the resistance, quality factor and resonance
frequency of the left and right reservoir, respectively. The
behavior of the system is now uniquely determined by nine
dimensionless parameters: $\epsilon/\omega_0$,
$k_BT_{{\rm low}/{\rm high}}/\hbar\omega_0$, $M_{L/R}^2I_0^2/\hbar R_{L/R}$,
$Q_{L/R}$, and $\omega_{L/R}/\omega_0$. Rectification can then be
calculated from Eq.~(\ref{recti}), by computing the forward and
reverse bias currents, $J_{+/-}$, with the above prescription.

\section{Results}

\begin{figure}[h]
\centering
\includegraphics[width=\columnwidth,clip]{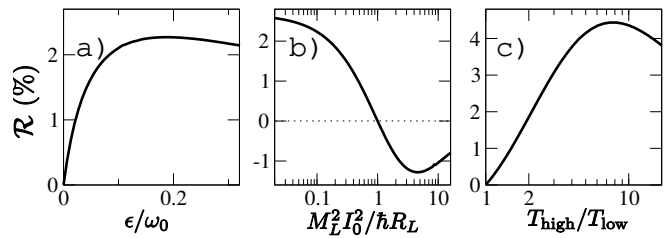}
\caption{Rectification with purely resistive reservoirs, $Q_L=Q_R=0$,
as a function of (a) nonlinearity $\epsilon$, (b) coupling $M_L$, and
(c) temperature ratio $T_{\rm high}/T_{\rm low}$.
In all panels we have
$\epsilon/\omega_0=0.07$, $k_BT_{\rm high}/\hbar\omega_0=0.2$,
$k_BT_{\rm low}/\hbar\omega_0=0.1$, $M_{L}^2I_0^2/\hbar R_{L}=0.2$,
and $M_{R}^2I_0^2/\hbar R_{R}=1$, except for the variable on the horizontal
axis. In panel (c) $T_{\rm high}$ is varied.}\label{fig2}
\end{figure}

Let us illustrate some generic features of the model with the simple
setup of two purely dissipative reservoirs, $Q_L=Q_R=0$, in which
case the frequencies $\omega_L$ and $\omega_R$ are irrelevant. In
Fig.~\ref{fig2} we plot the rectification against three different variables.
First, from Fig.~\ref{fig2}(a) we see that already at quite small values of
nonlinearity, $\epsilon\sim 0.1\,\omega_0$, the rectification has
essentially reached its maximum. Such values for $\epsilon$
are well within the regime of validity of our approximations and
should also be easily achieved in the experimental setup proposed
below. Next, Fig.~\ref{fig2}(b) exemplifies a very generic feature: having
$M_L^2/R_L < M_R^2/R_R$ tends to produce $J_+>J_-$, and vice versa.
Finally Fig.~\ref{fig2}(c) shows that the rectification increases
logarithmically with the temperature ratio $T_{\rm high}/T_{\rm low}$. Therefore, to
see an appreciable effect, the temperature difference $T_{\rm high}-T_{\rm low}$
should be of the same order of magnitude as the temperatures
themselves.

\begin{figure}[h]
\centering
\includegraphics[width=\columnwidth,clip]{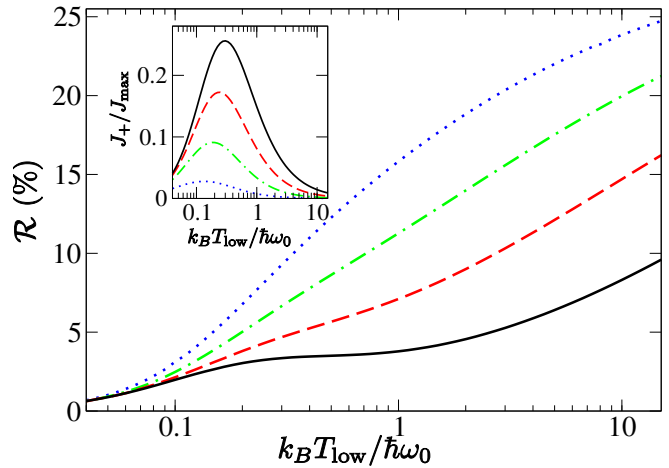}
\caption{Rectification with one reactive reservoir ($Q_L=0.1$). Here
$T_{\rm high}/T_{\rm low}=2$ and the different curves correspond to
$\omega_L/\omega_0=0.2$ (solid), 0.1 (dashed), 0.05 (dash-dotted),
0.02 (dotted). Other parameters as in Fig.~\ref{fig2}.}\label{fig3}
\end{figure}

For purely resistive reservoirs maximal value for the rectification
is about 2\% (Fig.~\ref{fig2}(a)). Larger values can be obtained by adding a
reactive part to one of the reservoir circuits. Then, as Fig.~\ref{fig3}
shows, $\cal R$ can be made an order of magnitude higher.  The inset
shows the current $J_+$, normalized with respect to the universal
single-channel maximum heat current $J_{\rm max}=\frac{\pi
k_B^2}{3\hbar}(T_{\rm high}^2-T_{\rm low}^2)$. \cite{pendry}
According to Fig.~\ref{fig3}, the
highest values for $\cal R$ are obtained for high temperatures,
where $J_+$ tends to zero.  High rectification and large current are
thus competing effects, and the optimal operating point depends on
the experimental constraints. In any case, it is possible to obtain
a rectification of $\sim5\%$ with $J\sim0.1\,J_{\rm max}$ and up to
$\sim15\%$ with $J\sim0.01\,J_{\rm max}$.

\begin{figure}[t]
\centering
\includegraphics[width=\columnwidth,clip]{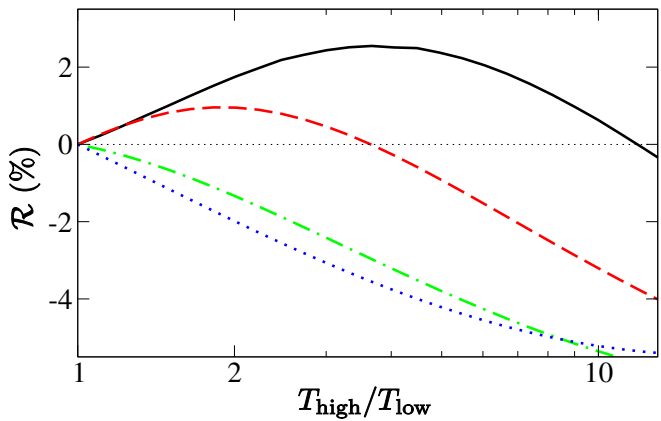}
\caption{Change of rectification sign as a function of operating
temperature. Here the lower temperature is kept fixed,
with values $k_BT_{\rm low}/\hbar\omega_0=0.1$ (solid),
0.3 (dashed), 1 (dash-dotted), 3 (dotted),
and the higher temperature is varied. The left reservoir is reactive
with $Q_L=0.1$ and
$\omega_L=10\,\omega_0$, and other parameters are
as in Fig.~\ref{fig2}.}\label{fig4}
\end{figure}

From Fig.~\ref{fig3} we also see that decreasing $\omega_L$ increases
rectification, so both small $\omega_L$ and the condition $M_L^2/R_L
< M_R^2/R_R$ favor the direction $J_+>J_-$. We can also combine
these two trends in an opposing manner by making $\omega_L$ large.
This way one can produce a system where the \emph{direction} of
rectification changes as a function of temperature.
From Fig.~\ref{fig4}
we see that in a system with a high-frequency reservoir (here $\omega_L=10
\,\omega_0$), ${\cal R}$ is positive when both temperatures
are below $\hbar\omega_0/k_B$, but at higher
temperatures the same device produces a negative ${\cal R}$.
In contrast to previous reports\cite{hu,segal3} on the rectification
sign reversal, in our system only the reservoir temperatures need to be
changed, not the device parameters.

The direction of rectification can be understood as follows.
Equations (\ref{iiless}) and (\ref{phi}) show that
the mean field $\Phi$ is proportional to the self-energy $\Sigma^<$.
Due to the self-consistency loop the relationship between
$\Phi$ and $\Sigma^<$ is not actually linear, but in practice making
$\Sigma^<$ larger will also increase $\Phi$.
Next, Eq.~(\ref{sigmaless})
shows that $\Sigma^<$ is essentially the product of Bose
function $n_{L/R}(\omega)$ and the effective coupling strength
$\omega M_{L/R}^2{\rm Re}[Y_{L/R}({\omega})]$, summed over the
two reservoirs. Because of this form, it follows that when
comparing the forward and reverse bias settings, larger $\Sigma^<$
is obtained in the case when the more strongly coupled reservoir is hotter.
As a consequence, the mean field $\Phi$ is also larger when
the more strongly coupled reservoir is hotter.

To interpret physically 
the $\Phi$-dependence of the current $J$, we analyze separately the
numerator and denominator of the integrand in Eq.~(\ref{curr5}).
The numerator is the product of the energy
$\hbar\omega$, Bose window $n_L(\omega)-n_R(\omega)$, and
effective reservoir coupling strengths, as defined above. Thus it
can be seen as a measure of the energy available for transport at the reservoirs.
On the other hand, the denominator is due to the Green's function
$|\langle\hat{I}\hat{I}\rangle^{r}(\omega)|^2$, giving the
transmittance of the central circuit. Equation (\ref{curr5}) shows
that an increasing $\Phi$ effectively increases the central circuit
resonance frequency, thereby shifting the resonator transmission window
to higher energies. Because of the Bose functions, in most
situations the numerator
is smaller at higher energies and the total current decreases.
But this is not always the case. It turns out that
if at least one of the reservoirs is reactive with high resonance
frequency ($\gtrsim\omega_0$) and the reservoir temperatures are
high ($\gtrsim\hbar\omega_0/k_B$), the peak of the numerator
is shifted to high enough energies so that an increasing $\Phi$
produces on increasing $J$.
In summary, except for the case of high-temperature and high-frequency
reservoirs, larger current is obtained in the configuration where
the more weakly coupled reservoir is hotter. This explains the sign
of ${\cal R}$ in all our results.

\section{Experimental realization}

\begin{figure}[h]
\centering
\includegraphics[width=\columnwidth,clip]{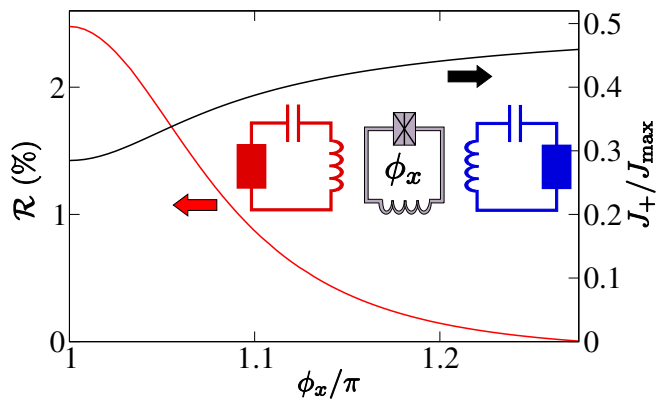}
\caption{Rectification and heat current through the SQUID
system as a function of the control flux $\phi_x$.
Parameter values at $\phi_x=\pi$ as in Fig.~\ref{fig2}, except
$T_{\rm high}=2\,T_{\rm low}=0.4\hbar\omega_0/k_B$.
Inset:
Schematic of the SQUID setup with two linear reservoir circuits
inductively coupled to a superconducting loop containing one Josephson junction.}
\label{fig5}
\end{figure}

For low operating temperatures, with $T_{\rm high}$, $T_{\rm low}$ approximately in
the range 100~mK--1~K, the studied model can be realized by the
setup shown in Fig.~\ref{fig5}. The system consists of a superconducting loop
containing a Josephson junction characterized by its Josephson
energy $E_J$ and shunt capacitance $C$. The loop itself is assumed
to have a finite inductance dominating the potential landscape. The
Hamiltonian of the system is  \cite{makhlin}
\begin{align}\label{effect}
H_M=E_C\hat{q}^2+E_L(\hat{\phi}-\phi_x)^2-E_J\cos\hat{\phi},
\end{align}
where the charging and inductive energies are $E_C=e^2/2C$,
$E_L=(\hbar/2e)^2/2L$, and $\phi_x$ denotes the external magnetic
flux through the loop (in units of $\hbar/2e$). The superconducting
phase across the junction $\hat{\phi}$ and the charge at the
capacitor $\hat{q}$ (in units of electron charge) are treated as
conjugate observables $[\hat{\phi},\hat{q}]=2i$. The charging term
can be thought of as the kinetic energy and the $\phi$-dependent
terms as an effective potential energy of a fictitious particle. In
the following we assume that $\phi_x\approx \pi$ and $E_J<2E_L$ so
that the potential has a single minimum at $\hat{\phi}=\phi_0$, with
$\phi_0\approx\pi$. With these assumptions  the phase is bound close
to the minimum so that we can approximate the potential accurately
by expanding the cosine term to the 4th order:
\begin{align}\label{effect1}
H_M&=E_C\hat{q}^2+(E_L+\frac 1 2 E_J\cos\phi_0)\hat{\phi}^2-
\frac{1}{24}E_J\cos\phi_0\,\hat{\phi}^4\nonumber\\
&\equiv E_C\hat{q}^2+E_2\hat{\phi}^2+E_4\hat{\phi}^4,
\end{align}
 the second line defining the quantities $E_2$ and $E_4$.
In general there should also be a $\hat{\phi}^3$ term, but with
$\phi_0\approx\pi$ this is small.  Further, within the mean-field
approximation one has
$\hat{\phi}^3\sim\hat{\phi}\langle\hat{\phi}^2\rangle$, producing
just a shift in the origin. Writing the charge and phase in terms of
bosonic creation and annihilation
operators we recover exactly Eq.~(\ref{nonlin}) 
with parameters $\hbar\omega_0 = 4\sqrt{E_CE_2}$ and $\hbar\epsilon
= 2\frac{E_CE_4}{E_2}$. The current operator of the circuit is given
by $\hat{I}=I_0(\hat{b}+\hat{b}^\dagger)$, where
$I_0=4e(E_CE_2^3)^{1/4}/\hbar$. Thus, in the parameter regime
$E_2\gg E_4$ we have effectively realized our
weakly nonlinear resonator model.

As the above considerations show, varying the externally applied
field $\phi_x$ about $\pi$ moves the potential minimum $\phi_0$
which in turn changes the values of the parameters $\omega_0$ and
$\epsilon$. In particular, $\epsilon$ is maximized at
$\phi_x=\phi_0=\pi$ and vanishes when $\phi_0\to\pi\pm\pi/2$. Figure~\ref{fig5}
demonstrates the resulting continuous tuning of rectification
performance.

\section{Conclusions}

In conclusion, we have analyzed heat rectification effects in
radiative heat transport through a nonlinear quantum resonator. This
system is particularly interesting because it can be realized by an
experimentally feasible superconducting circuit. The proposed system
is operated in a low-temperature regime and can be controlled by 
by applying external magnetic fields.
Despite its simplicity, the system is capable of producing a rectification
of over 10\%. We have shown that in a suitable parameter regime the
direction of rectification changes as a function of temperature and given a physical explanation for the phenomena.


\begin{thebibliography}{99}
\bibitem{schwab}K. Schwab, E.~A. Henriksen, J.~M. Worlock, and M.~L. Roukes, Nature (London) {\bf 404}, 974 (2000).
\bibitem{meschke} M. Meschke, W. Guichard, and J.~P. Pekola, Nature {\bf 444}, 187 (2006).
\bibitem{saira}O.-P. Saira \emph{et al.}, Phys. Rev. Lett. {\bf 99}, 027203 (2007).
\bibitem{schmidt}D.~R. Schmidt, R.~J. Schoelkopf, and A.~N. Cleland, Phys. Rev. Lett. {\bf 93}, 045901 (2004).
\bibitem{ojanen}T. Ojanen and  T.~T. Heikkil\"a, Phys. Rev. B {\bf 76}, 073414 (2007).
\bibitem{terraneo}M. Terraneo, M. Peyrard and G. Casati , Phys. Rev. Lett. {\bf 88}, 094302 (2002).
\bibitem{li}B. Li, L. Wang and G. Casati, Phys. Rev. Lett. {\bf 93}, 184301 (2004).
\bibitem{hu}Bambi Hu, Lei Yang, and Yong Zhang, Phys. Rev. Lett. {\bf 97}, 124302 (2006).
\bibitem{segal2}D. Segal and A. Nitzan, Phys. Rev. Lett. {\bf 94}, 034301 (2005).
\bibitem{zeng}N. Zeng and J.-S. Wang, Phys. Rev. B {\bf 78}, 024305 (2008).
\bibitem{segal}D. Segal, Phys. Rev. Lett. {\bf 100}, 105901 (2008).
\bibitem{wu}L.-A. Wu and D. Segal, Phys. Rev. Lett. {\bf 102}, 095503 (2009).
\bibitem{chang}C. W. Chang, D. Okawa, A. Majumdar and A. Zettl, Science {\bf 314}, 1121 (2006).
\bibitem{scheibner}R.~Scheibner, M.~K\"{o}nig, D.~Reuter, A.~D.~Wieck, C.~Gould, H.~Buhmann, and L.~W.~Molenkamp,
New J. Phys. {\bf 10}, 083016 (2008).
\bibitem{ojanen2}T. Ojanen and A.-P Jauho, Phys. Rev. Lett. {\bf 100}, 155902 (2008).
\bibitem{remark} In Eq.~(\ref{current4})  heat current is positive when energy flows from left to right. Equation~(13) in Ref.~\onlinecite{ojanen2} has an erroneous sign.
\bibitem{pendry}J.~B. Pendry, J. Phys. A: Math. Gen.  {\bf 16}, 2161 (1983).
\bibitem{haug} H. Haug and A.-P. Jauho, \emph{Quantum Kinetics in Transport and Optics of Semiconductors} (Springer-Verlag, Berlin Heidelberg, 1996).
\bibitem{segal3}D. Segal and A. Nitzan, J. Chem. Phys. {\bf 122}, 194704 (2005).
\bibitem{makhlin} Yu. Makhlin, G. Sch\"on, A. Shnirman, Rev. Mod. Phys. {\bf 73},
357 (1999).

\end{thebibliography}
\end{document}